\documentclass[aps,pre,twocolumn,showpacs]{revtex4}
\usepackage{epsfig}
\usepackage{times}
\begin{document}

\title{Motion of influential players can support cooperation in Prisoner's Dilemma}

\author{Michel Droz$^\star$, Janusz Szwabi{\'n}ski$^{\star\dagger}$ and Gy{\"o}rgy Szab{\'o}$^\ddagger$}
\affiliation {$^\star$Department of Physics, University of Geneva,
 CH-1211 Geneve 4, Switzerland\\
$\dagger$ Institute of Theoretical Physics, University of Wroc{\l}aw,\\
pl. M. Borna 9, 50-204 Wroc{\l}aw, Poland
\\
$^\ddagger$Research Institute for Technical Physics and Materials
Science, P.O. Box 49, H-1525 Budapest, Hungary}

\begin{abstract}
We study a spatial Prisoner's dilemma game with two types ($A$ and $B$) of players located on a square lattice. Players following either cooperator or defector strategies play Prisoner's Dilemma games with their 24 nearest neighbors. The players are allowed to adopt one of their neighbor's strategy with a probability dependent on the payoff difference and type of the given neighbor. Players $A$ and $B$ have different efficiency in the transfer of their own strategy therefore the strategy adoption probability is reduced by a multiplicative factor ($w < 1$) from the players of type $B$. We report that the motion of the influential payers (type $A$) can improve remarkably the maintenance of cooperation even for their low densities.
\end{abstract}


\maketitle

\section{Introduction}
\label{sec:introduction}

For the consideration of cooperation among selfish individuals the application of evolutionary Prisoner's Dilemma (PD) games proved to be a fruitful mathematical background \cite{weibull_95}. In the original two-person one-shot game the equivalent players have two options [to cooperate ($C$) or to defect ($D$)] to choose and their payoffs depend on their choices. The highest total payoff is achieved and shared equally if both player choose $C$. On the contrary, the players share equally the lowest total income when both choose defection. The highest individual payoff is received by the defector against the cooperator co-player who obtains the lowest individual payoff. The selfish players are enforced to choose defection that yields better score for any choice of the co-player. In the traditional game theory the players are intelligent, thus both selfish individual choose defection providing the second lowest income for the players.

During the last decades the original concepts of game theory \cite{neumann_44} were extended for different directions that includes the introduction of uncertainties, multi-agent repeated games, evolutionary rules, {\it etc}. Due to the progressive research many ways were discovered how the cooperation can be maintained in a society of selfish individuals as it is observed in real biological \cite{maynard_82,kerr_n02} and human systems \cite{axelrod_84}. The most relevant mechanisms supporting cooperation are the kin selection \cite{hamilton_jtb64a}, direct reciprocity \cite{trivers_qrb71,axelrod_84}, indirect reciprocity \cite{nowak_n98,fehr_n02}, group selection \cite{traulsen_pnas06},
and spatial systems with short range interactions between the players \cite{nowak_ijbc93} (for comparison and further references consult the paper by Nowak \cite{nowak_s06}).

In the spatial evolutionary PD games the players' payoff come from
games with their neighbors and the players can adopt a strategy
from one of their neighbors with a probability dependent on the
payoff difference. Most of the early works were concentrated on
the evaluation of the average density of cooperators when varying
the model parameters, like the set of strategies, the evolutionary
rules including noises, the payoff values, and the structure of
connectivity (for a survey see \cite{nowak_06,szabo_pr07}). It turned out
that cooperators cannot remain alive in the spatial evolutionary PD games if the temptation to choose defection (defector's income against cooperator) exceeds a threshold value dependent on the mentioned parameters. In contrary to spatial evolutionary PD games, more than 80 percent of payers choose cooperation within the whole range of payoff parameters in the models suggested by Santos {\it et al.} 
\cite{santos_prl05,santos_prslb06} where the players were located on the sites of a scale-free network. 

For the investigation of human societies the so-called social networks provide a more appropriate connectivity structure and different versions of evolutionary PD games were studied on small-world, scale-free, and other networks too \cite{abramson_pre01,ebel_pre02,masuda_pla03}. An extremely large enhancement in the portion of cooperators is occurring when the evolutionary rule is controlled by the difference of total incomes
that favors the strategy adoption from those players who have a large number of neighbors \cite{santos_prl05,santos_prslb06}. In these models the players with many neighbors play a crucial role in the maintenance of cooperation because their strategy is adopted by their neighborhood and this process is beneficial for cooperators while it decreases the defector's income. The same mechanism can occur and support cooperation for those models where a portion of players have enhanced activity in spreading their own strategy over their neighborhood \cite{szolnoki_epl07}. In real human societies these latter players can represent influential players and masters as well as prophets or agitators.

Some enhancement in the density of cooperators was already reported by several authors who considered the effect of inhomogeneous strategy adoption probabilities \cite{kim_pre02,wu_cpl06,wu_pre06}. The most significant increase of the cooperative behavior is found for those types of inhomogeneities where each player is characterized by a strategy transfer capability quantifying the probability of strategy adoption from the given player to her neighbors \cite{szolnoki_epl07,szolnoki_epjb08}. Subsequent investigations have clarified that the efficiency of this
mechanism can be improved if the number of neighbors is increased
even for regular connectivity structures \cite{szabo_arx08}. Furthermore, it turned out that the co-evolution of strategy distribution and strategy transfer capability yields an inhomogeneity in the strategy transfer activity supporting the cooperative behavior \cite{szolnoki_njp08}.

In the present paper the above investigations are extended to study what happens in a spatial evolutionary PD games if the number of neighbors is large (24) while the density of influential players is low. Such a large neighborhood is natural in social systems. Besides it, the large number of neighbors enhances the phenomenon and allow us to visualize its main features. For low densities of influential players the direct links between these players are rare. Consequently, the cooperative behavior cannot spread away through these direct connections. Now we show that this difficulty can be overcome if the influential players are allowed to migrate. In this case the temporary connections between the latter players can provide suitable conditions for the cooperators to rule over the whole system.

\section{The model}
\label{sec:model}

We consider an evolutionary two-strategy Prisoner's Dilemma game with players located on the sites ($x$) of a square lattice. Two types of players are distinguished and their spatial distribution is described by an Ising formalism ($n_x=A$ or $B$). The portion of players $A$ and $B$ are fixed [$\nu$ and $(1-\nu)$]. The player at site $x$ can follow either an unconditional cooperator (${\bf s}_x=C$) or defector (${\bf s}_x=D$) strategy, denoted also by unit vectors as
\begin{equation}
\label{eq:cd} {\bf s}_x=C= \left( \matrix{1 \cr 0 \cr }\right)
\;\; \mbox{or}\;\; D= \left( \matrix{0 \cr 1 \cr }\right) \;.
\end{equation}
This notation allows us to use a simple matrix algebra for the definition of the total income $U_x$ of player $x$ coming from PD games played with her all neighbors $y \in \Omega_x$, that is,
\begin{equation}
\label{eq:tpo} U_x=\sum_{y \in \Omega_x } {\bf s}^{+}_x {\bf A}
\cdot {\bf s}_y\,\,,
\end{equation}
where ${\bf s}^{+}_x$ is the transpose of the state vector ${\bf s}_x$, and the summation runs over all the neighbors of player $x$. In the present case each player has 24 neighbors ($|\Omega_x|=24$) located inside a block of $5 \times 5$ sites around the central player $x$. Following the notation suggested by Nowak {\it et al.} \cite{nowak_ijbc93} we use the rescaled payoff matrix:
\begin{equation}
%\label{eq:pom}
{\bf A}=\left( \matrix{1 & 0 \cr
                       b & 0 \cr} \right)\;, \;\; 1 < b < 2\,,
\end{equation}
where we have only one parameter $b$ characterizing the temptation to choose defection. In the present evolutionary PD game a randomly chosen player $x$ could adopt the strategy from one of its randomly chosen neighbors $y \in \Omega_x$ with a probability $W[({\bf s}_x \to {\bf s}_y)$ depending on the payoff difference and the type of player $y$
\cite{szolnoki_epl07}. Namely,
\begin{equation}
\label{eq:update} W({\bf s}_x \to {\bf s}_y) = w_y {1 \over 1 +
 \exp {[(U_x-U_y)/K]} } \;,
\end{equation}
where $K$ characterizes the uncertainties (stochastic noises) in the value of total payoff \cite{perc_njp06a,perc_pre07,traulsen_jtb07} and/or a freedom for the players to make irrational decisions when adopting a strategy \cite{blume_geb93,szabo_pre02d}. The multiplicative factor $w_y$ defines the strategy transfer capability of the player $y$ in a way that
\begin{equation}
\label{eq:wobt} w_{y}= \cases {1, &if $n_y=A$ \cr
                w, &if $n_y=B$ \cr }\; .
\end{equation}
Players of type $A$ are considered as influential players who are capable to convince their neighbors (with a high efficiency on comparison to players of type $B$) to follow them in the choice of strategy.

The system started from a state where a fraction $\nu$ of players (distributed randomly) belong to the type $A$ and the rest of players are $B$. In the random initial state the players follow $C$ or $D$ strategies with equal probability independently of their types. During one Monte Carlo step (MCS) each player has a chance once on average to adopt a strategy from one of their neighbors (chosen at random) as described above. Besides it, the influential players are allowed to move. More precisely, after each MCS a fraction $f$ of $A$ players (chosen at random) can exchange their site with one of the randomly selected nearest neighbors $y$ if $n_y=B$, that is, $({\bf s}_x,{\bf s}_y) \to ({\bf s}_y,{\bf s}_x)$, $n_x=A \to B$, and $n_y=B \to A$. The magnitude of $f$ characterizes the migration (diffusivity) of influential players. The simulations were performed on a square lattice with a size $L \times L$ under periodic conditions. After a suitable thermalization time $t_t$ we have evaluated the concentration $\rho$ of cooperators in the stationary states by averaging over a sampling time $t_s$. Most of the MC simulations were performed for $L=400$, $K=2.4$, and $\nu=0.02$ for different values of $w$, $b$, and $f$. As the relaxation (thermalization) time depends on $w$ therefore $t_t=t_s$ is varied from $10^4$ to $10^6$ MCS. The longer run time is used for small values of $w$ when most of the players (of type $B$) modify her strategy with a low frequency proportional to $w$.

\section{Monte Carlo results for quenched distribution of types}
\label{sec:qres}

\begin{figure}[ht]
\centerline{\epsfig{file=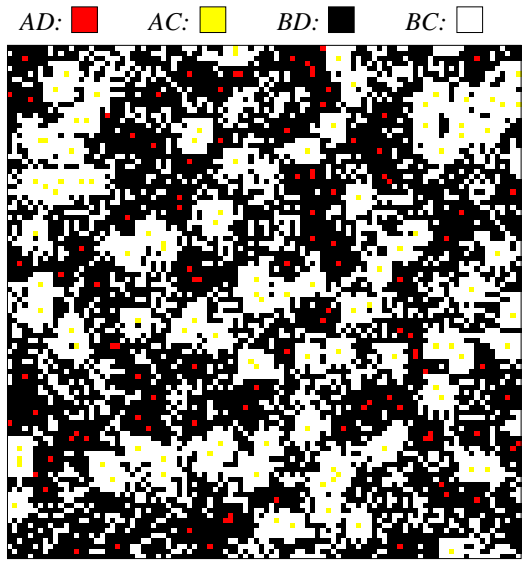,width=7cm}}
\caption{\label{fig1}(Color online) A typical distribution of the strategies ($D$ or $C$) and types ($A$ or $B$) of players on a square lattice (with a block size of $100 \times 100$) for a quenched random distribution of players $A$ if $\nu=0.02$, $b=1.25$, $K=2.4$, and $w=0.001$.}
\end{figure}

First we investigate the system when the motion of players $A$ is forbidden ($f=0$) for a small density ($\nu= 0.02$) which is significantly lower than the optimum ($\nu_{\rm opt} \simeq 0.2$) discussed in a previous paper \cite{szabo_arx08}. Figure \ref{fig1} illustrates the main features of the spatial distribution of strategies and types ($s_x$ and $n_x$) for a low value of $w$. The snapshot shows clearly that players of type $A$ are surrounded by players following the same strategy. This means that the income of cooperating $A$s (in short, $AC$ players) is enhanced by their neighborhood ($BC$ players) while the defecting $A$s receive a very low payoff. In fact, this short range correlation (in the strategy distribution) is the reason why cooperators can survive for the given parameters ($b$ and $K$) ensuring survival only for defectors in the homogeneous system ({\it i.e.} for $\nu =0$ or $1$, or for $w=1$ at arbitrary $\nu$). 

For such a large neighborhood and a large value of $1/w$ one can think that the short-time dynamics (strategy adoption) between two neighboring $A$ players can be approximated by introducing an effective payoff matrix as it was described by Pacheco {\it et al.} \cite{pacheco_prl06} when studying the co-evolution of strategy distribution and connectivity structure.
One can observe in Fig.~\ref{fig1} that both the $AC$ and $AD$ players form small colonies. We have to emphasize that the strategy distribution varies continuously and due to the stochastic noise even an $AD$ player can be transformed into $AC$ and {\it vice versa}. For the present low value of $\nu$ the overlapping neighborhood of the $A$ players do not span the whole system. Consequently, the strategy fluctuations in the intermediate regions play a crucial role in the coexistence of $D$ and $C$ strategies.

For the quantitative analysis MC simulations were performed to determine the average density ($\rho$) of cooperators when varying the payoff parameter (temptation $b$) for several values of $w$ while other parameters are fixed. 
\begin{figure}[ht]
\centerline{\epsfig{file=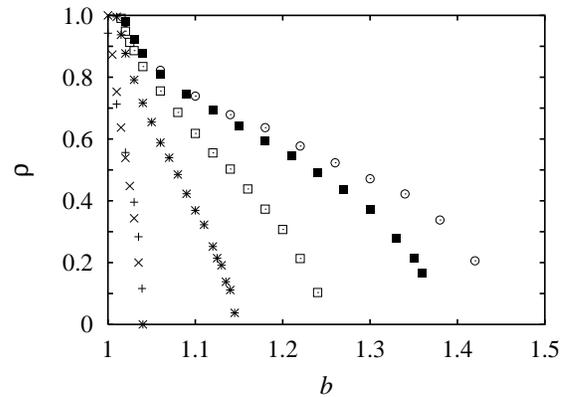,width=8cm}}
\caption{\label{fig2}Average density of cooperators in the stationary state (within the coexistence region) as a function of $b$ for six different values of $w$ (from left to right $w=1$, 0.2, 0.05, 0.02, 0.005, and 0.002) if $K=2.4$ and $\nu=0.02$.}
\end{figure}
Figure \ref{fig2} compares six curves describing a monotonous decrease of $\rho$ from 1 to 0 within a coexistence region where $0 < \rho < 1$. This means that only defectors remain alive if $b_{c1}(K,w,\nu) < b$ and cooperators prevail the whole systems if $ b < b_{c2}(K,w,\nu)$ (both threshold values depend on the model parameters). The effect of inhomogeneous teaching activity is practically negligible if the ratio $1/w$ is not large enough. Notice that two curves (obtained for $w=1$ and $0.2$) practically coincide in Fig.~\ref{fig2}. On the contrary, the density of cooperators as well as the second threshold value of temptation ($b_{c2}(K,w,\nu)$) is increased significantly if $w$ becomes very small. At the same time, the MC results indicate only a small increase in the first threshold value of temptation [$b_{c1}(K,w,\nu)$]. 

If parameters are tuned in the homogeneous ($w=1$) spatial system, then the extinction of cooperators (or even defectors) exhibits a critical phase transition belonging to the directed percolation universality class \cite{janssen_zpb81,grassberger_zpb82,szabo_pre98}. This means that the decrease of density follows a power law behavior when approaching the critical point, that is $\rho \sim |b-b_c|^{\beta}$ if $b \to b_c$, where the value of exponent $\beta$ is determined by the spatial dimension. The algebraic decrease of $\rho$ is accompanied by a divergency in fluctuations, correlation length, and relaxation time (for details see \cite{marro_99,hinrichsen_ap00}). These features cause technical difficulties in the accurate determination of $\rho$ because long run time and large system size are required in the close vicinity of the critical point. The technical difficulties become more pronounced if a Griffiths-like phase occur in the inhomogeneous spatial system \cite{griffiths_prl69}. This is the reason why data with a small value of $\rho$ are missing in Fig.~\ref{fig2} for low values of $w$.

\begin{figure}[ht]
\centerline{\epsfig{file=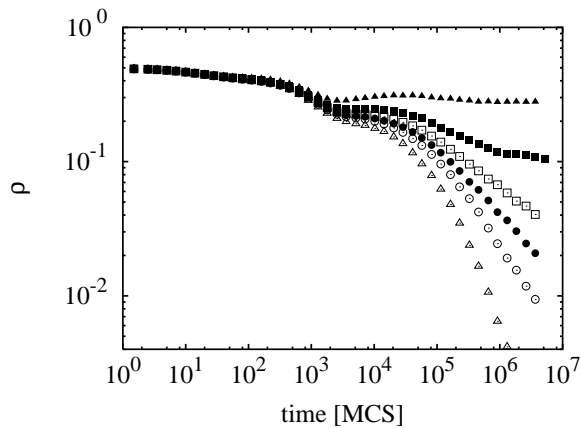,width=8cm}}
\caption{\label{fig3}Log-log plot of the time-dependence of the density of cooperators for different values of $b$ (from top to bottom $b=1.4$, 1.45, 1.47, 1.48, 1.49, and 1.51) if $w=0.002$, $K=2.4$, $\nu=0.02$, and $L=1800$.}
\end{figure}

Figure \ref{fig3} show several examples about how the density $\rho(t)$ of cooperators evolves towards the final stationary state in the vicinity of the second transition point. Despite of the large system size ($L=1800$) the vanishing density of cooperators exhibits some relevant fluctuation preventing the clear visualization of the average behavior in the log-log plot for sufficiently long times. In order to suppress the undesired disturbance of fluctuation the data in Fig.~\ref{fig3} are averaged over a time window ($0.8 t_n < t < 1.2 t_n$ where $t_n \simeq 2^{n/2}$) with an interval increasing linearly with $t$. The smoothed data show clearly that the density of cooperators decrease algebraically ($\rho(t) \sim t^{-\delta}$) with an exponent $\delta > 0$ dependent on $b$. The upper curve of this plot illustrates an example where $\rho(t)$ tends to a finite limit value. 

Similar behavior was reported for other simpler models (e.g., contact process) when considering the extinction of a species (or any other objects or states) on a quenched inhomogeneous spatial background \cite{noest_prl86,noest_prb88,moreira_pre96,dickman_pre98,szabo_pre02b}.  
On the inhomogeneous backgrounds we can distinguish patches providing better conditions for the species (henceforth strategies) to survive. For low densities ($\rho$) of the disappearing strategies the active territories are separated from each other and the whole process can be well approximated by the statistical description of independent extinctions on patches of different sizes. The average life-time increases with the size $s$ of the mentioned patches while the probability of their appearance decreases exponentially with $s$. Noest \cite{noest_prl86,noest_prb88} has shown that the resulting process yields an algebraic decay. Recent theoretical investigations of the random contact process \cite{hooyberghs_prl03,hooyberghs_pre04,vojta_pre05,hoyos_pre08} are focused on clarification of phenomenon what happens when varying the strength of inhomogeneity (for a survey see \cite{igloi_pr05}). Similar behavior is expected in the present model. Unfortunately the numerical confirmation of the mentioned feature exceeds our computational capabilities. We have to emphasize, however, that technical difficulties are reduced if the inhomogeneous background changes continuously. In the latter case the system becomes equivalent to the homogeneous cases for sufficiently large time- and length-scales \cite{hinrichsen_ap00,szabo_pre02b} and this feature simplifies the numerical analysis as discussed below.

\section{Monte Carlo results for moving influential players}
\label{sec:mres}

\begin{figure}[ht]
\centerline{\epsfig{file=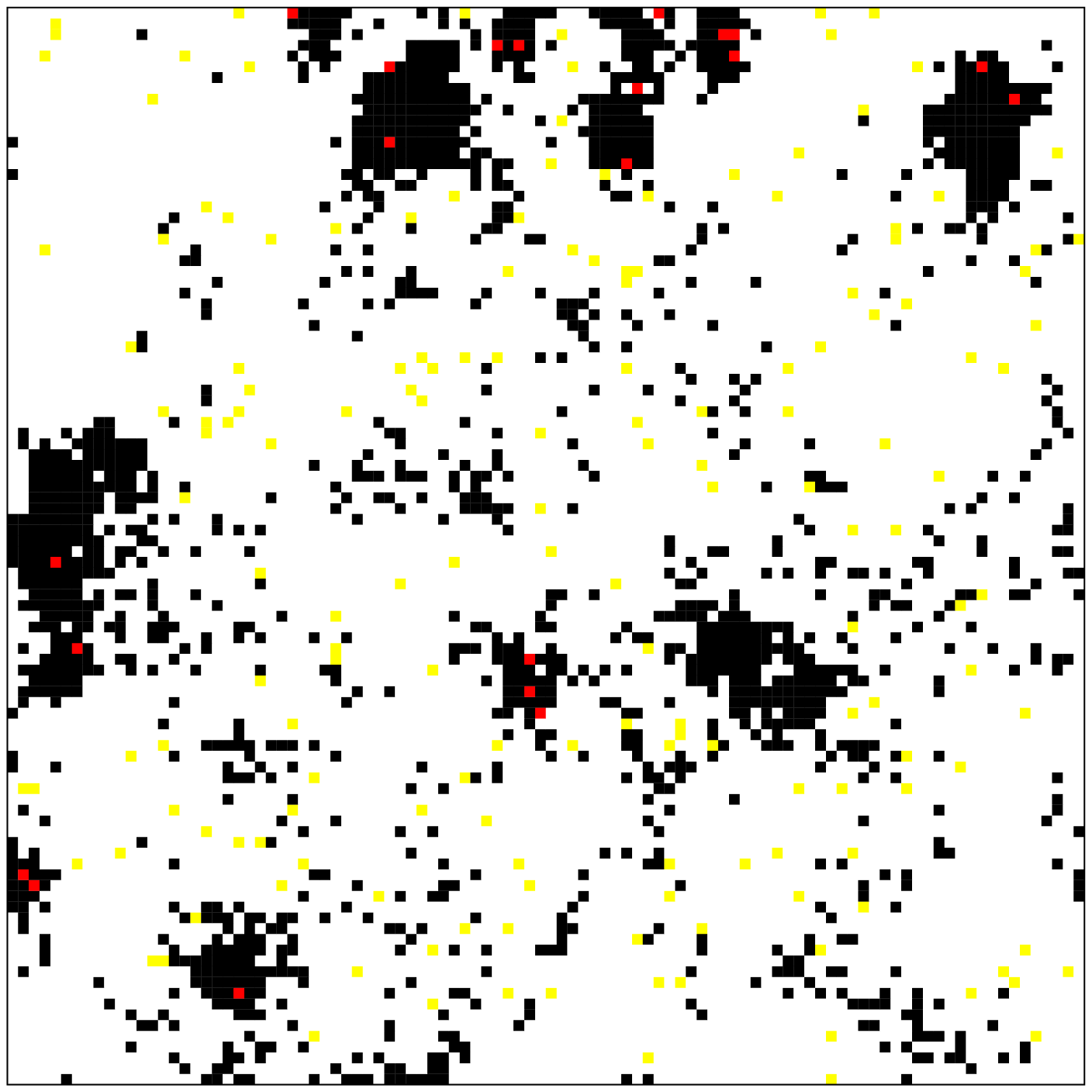,width=7cm}}
\centerline{\epsfig{file=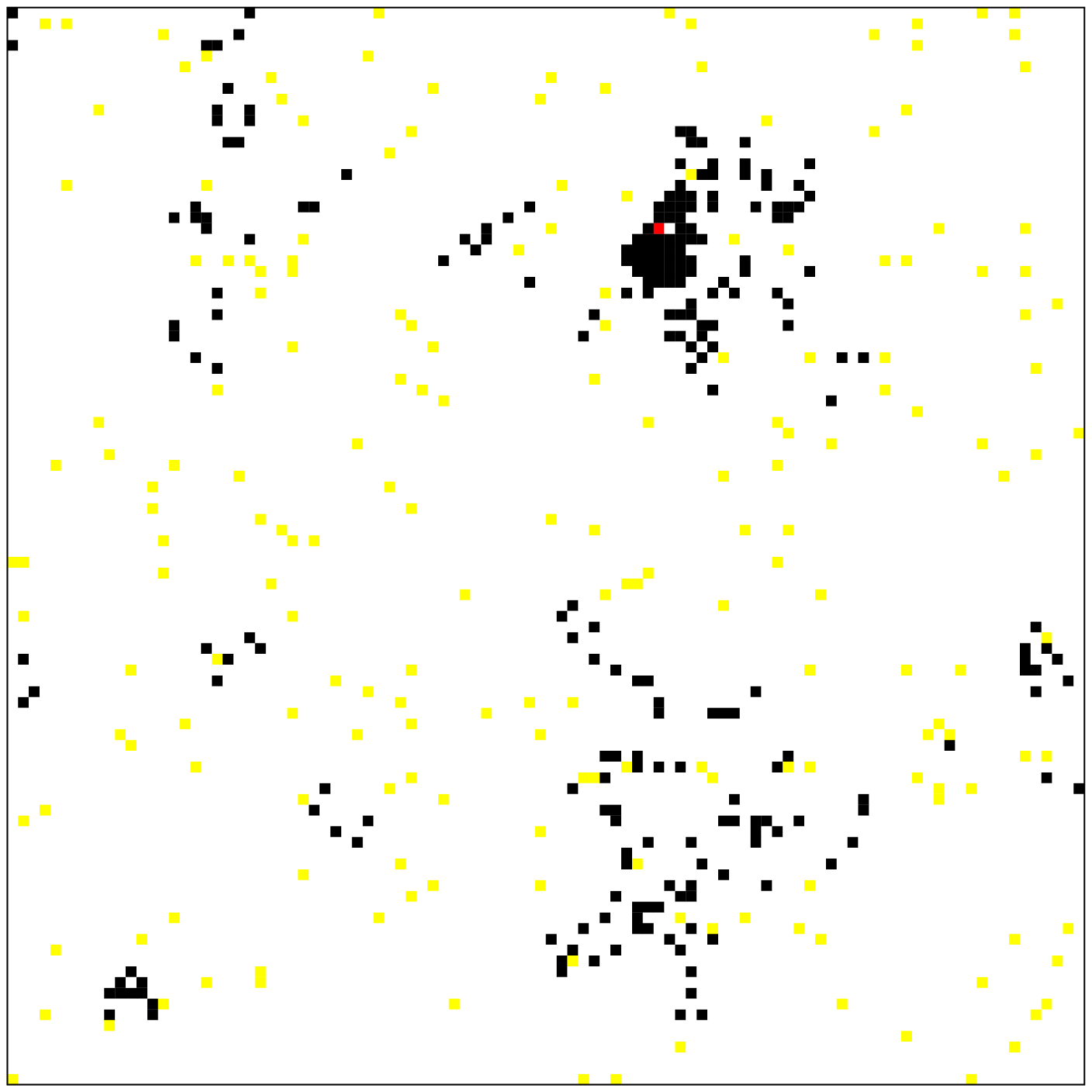,width=7cm}}
\caption{\label{fig4}(Color online) Two snapshots on the distribution of the strategies and types of players at times $t=1000$ MCS (upper plot) and $t=10000$ MCS (lower) for $\nu=0.02$, $b=1.25$, $K=2.4$, $f=0.1$, and $w=0.001$. In the final stationary state all the players cooperate. Notation of colors as in Fig.~\ref{fig1}}
\end{figure}

In this section we study the system when a slow motion of $A$ players is introduced. Most of the subsequent MC data are obtained for $\nu=0.02$ when 10 \% of players $A$ ($f=0.1$) are allowed to exchange her position with one of the neighbors as described above. In agreement with the expectations, for such a slow migration the $AC$ ($AD$) players are surrounded by cooperating $BC$s (defecting $BD$s). As a result, at the beginning of the evolutionary process one can observe a spatial distribution similar to the one plotted in Fig.~\ref{fig1}. For slow motion the given neighborhoods accompany the (central) influential players. Due to their motion the rare $A$ players can approach each other and when two of them interact then $AC$ convinces $AD$ to cooperate with a high probability and within a short time this new strategy will be adopted by the neighbors, too. Consequently, the number of $AD$ players decreases gradually as demonstrated in the upper snapshot of Fig.~\ref{fig4}.

In the present model the highest individual income is received by a solitary defector because she exploit all her cooperating neighbors. So the strategy of the solitary defector of type $B$ can be transferred to her neighborhood unless this player adopt cooperation from a neighboring $AC$ player. The lower snapshot in Fig.~\ref{fig4} shows a situation when the moving $AC$ players eliminate the (small) groups of $AD$ players. Sometimes, however, the strategy of the solitary defector can be adopted even by a neighboring $AC$ player who will enforce her neighbors to form a gang of defectors as illustrated in the lower snapshot of Fig.~\ref{fig4}. The formation of the defector gang reduces the income of the focal $AD$ player who will be conquered by an $AC$ player opposing her sooner or later and finally the defection becomes extinct in the whole system.  

\begin{figure}[ht]
\centerline{\epsfig{file=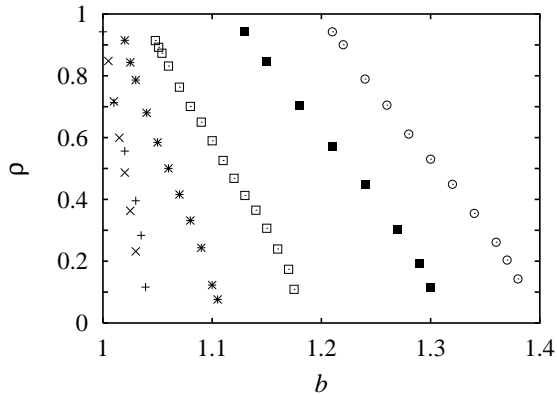,width=8cm}}
\caption{\label{fig5}Average density of cooperators versus $b$ if the motion of players $A$ is allowed. Parameters are the same as in Fig.~\ref{fig2} excepting that here $f=0.1$.}
\end{figure}

In order to quantify the efficiency of the above described mechanism we have determined the functions $\rho(b)$ at $f=0.1$. For the sake of comparison the rest of parameters are equivalent to those used in the previous section. The results plotted in Fig.~\ref{fig5} are similar to those obtained for quenched distributions of $A$ players at their higher densities (e.g., $\nu = 0.2$ \cite{szabo_arx08}). The most striking difference between the results of Figs.~\ref{fig2} and \ref{fig5} is that here the first transition occurs at higher values of $b$. In other words, the moving $AC$ players are capable to defeat those (rare) gangs of defectors which are stabilized for some quenched constellations. As well as for higher densities of $A$ players the coexistence region shifts towards the larger values of $b$ when the strategy transfer capability ($w$) of $B$ players is decreased. The results in Fig.~\ref{fig5} indicate a logarithmic increase in the critical values of temptation, i.e., $\delta b_c \sim \ln{1/w}$. The systematic analysis of this effect (for lower values of $w$) is prevented by the long relaxation time increasing with $1/w$. 

The visualization of the evolution of strategy distribution (for typical snapshots see Figs.~\ref{fig1} and \ref{fig4}) indicates clearly that the evolutionary process is mainly controlled by the competition between the moving $AC$ and $AD$ players surrounded by their own followers. In some sense the situation is analogous to the case of group (and/or kin) selection \cite{lehmann_pnas07,traulsen_pnas06,traulsen_bmb08,bowles_s06}. The fluctuating neighborhood of the moving $AC$ and $AD$ players induces uncertainties in the final results when they compete with each other. Besides it, the strategy distribution in the 'nobody territory' (consisting of sites not influenced by players $A$) can also affect the variation of strategy for players of type $A$. All these processes together yield a complex behavior dependent on the model parameters. In the next section the effect of mobility ($f$) is investigated quantitatively.

\section{Effect of mobility}
\label{sec:mobility}

In the limit of large mobility the advantage of the $AC$ players vanish because they cannot benefit from their followers left behind during their motion. Furthermore, their fast motion can be interpreted as a mixing favoring defection (see the results of mean-field approximation \cite{szabo_pr07}). In the opposite limit ($f \to 0$) the system is expected to reproduce a behavior discussed for the quenched distribution of $A$ players. Now we study what happens when varying the mobility of $A$ players.

\begin{figure}[ht]
\centerline{\epsfig{file=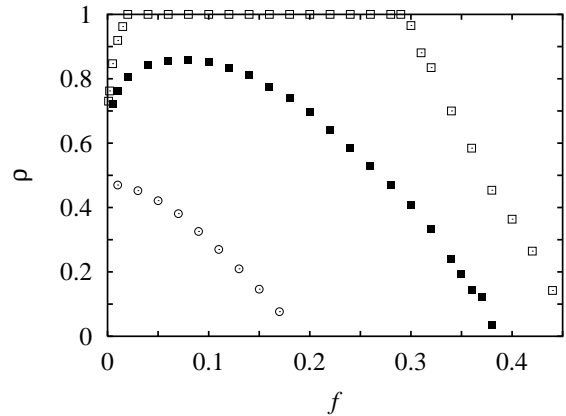,width=8cm}}
\caption{\label{fig6}Average density of cooperators as a function of mobility $f$ for $b=1.15$, $K = 2.4$, $\nu = 0.02$, and $w=0.002$, 0.005, and 0.02 (from top to bottom).}
\end{figure}

For low values of $1/w$ the small enhancement of cooperation is reduced gradually when $f$ is increased (see the lowest data obtained for w=0.02 in Fig.~\ref{fig6}). On the contrary, one can observe a local maximum in the density of cooperators at an optimum value of $f$ if $w=0.005$. The density of cooperators reaches its saturation value ($\rho=1$) within a suitable range of $f$ if $1/w$ exceeds a threshold value. For all the three plotted curves the density of cooperators vanishes if $f$ exceeds a threshold value dependent on the parameters.

\section{Effect of density of influential players}
\label{sec:mobility}

Previous investigations \cite{santos_prl05,santos_prslb06,rong_pre07} have indicated clearly that on the scale-free graphs the introduction of additional links between the influential players can suppress the mechanism supporting the emergence of cooperative behavior. Similarly, an optimal density $\nu$ of $A$ players was found on the two-dimensional lattices for quenched distribution of players $A$ and $B$. It turned out that the optimal value of $\nu$ depends mainly on the number of neighbors but it is also affected by other parameters (e.g., $b$ and $K$). 

\begin{figure}[ht]
\centerline{\epsfig{file=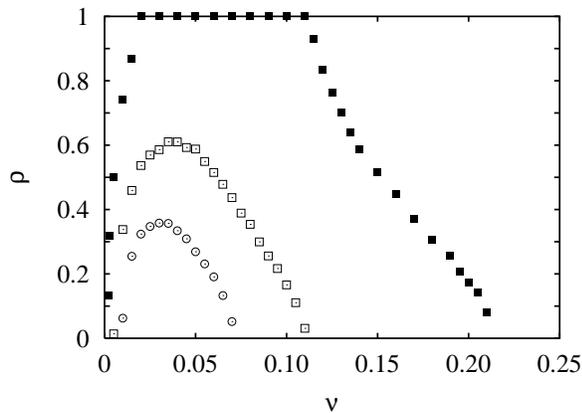,width=8cm}}
\caption{\label{fig7}Average density of cooperators {\it vs}. density $\nu$ of influential players for different values of temptation ($b=1.2$, 1.3, and 1.35 from top to bottom) at $f=0.1$, $K = 2.4$, and $w=0.002$.}
\end{figure}

Figure \ref{fig7} summarizes the results of MC simulations obtained for three different values of $b$ while other parameters are fixed. The results indicate clearly that optimal density of influential players is reduced particularly for such values of $b$ and $K$ where the cooperation can be maintained at a low level. The lowest curve in Fig.~\ref{fig7} shows clearly the appearance of a maximum at $\nu =\nu_{\rm opt} \simeq 0.03$ 
when varying $\nu$ if $K=2.4$, $b=1.35$, $f=0.1$, and $w=0.002$ while for these fixed parameters the cooperators remain alive only within a range of $\nu$. Similar behavior can be observed when the survival of $C$ strategy is supported by decreasing the temptation $b$. More precisely, the profile of the curve $\rho(\nu)$ becomes wider and higher until reaching the saturation value. Notice that the cooperators die out for all the three cases plotted in Fig.~\ref{fig7} if the density of influential players exceeds a value ($0.21$) close to the optimum for quenched disorder.

\section{Summary}
\label{sec:summary}

Within the framework of evolutionary Prisoner's Dilemma games we have studied the improvement of cooperative behavior with two types of players (both are following either cooperation or defection unconditionally) if the influential players (type $A$) are allowed to walk randomly through the whole square lattice. Our analysis is concentrated to systems with a small portion of influential players where the players have large neighborhood ($n=24$). In these cases the the influential players and their followers form an apparent group if there is a relevant difference between the strategy transfer capability between the $A$ and $B$ players. As a result, the evolution of strategy distribution is governed basically by the competition between the cooperative and defective influential players in such a way that the direct PD interaction (payoff) can be replaced by an effective interaction related to games with re-scaled payoffs. Similar phenomenon was described by Pacheco {\it et al}. \cite{pacheco_prl06,pacheco_jtb06} who studied the co-evolution of strategy distribution and connectivity structure. Besides it, the processes in the present model are resembling the kin and/or group selections \cite{hamilton_jtb64a,traulsen_pnas06,lehmann_pnas07} supporting cooperation, too. In comparison with the mentioned models, here the randomly moving groups (influential players) interact temporarily (if they are sufficiently close to each other). In the present case the strategy adoption between the influential players are affected by the time-dependent structure of groups and also by the strategy fluctuations in the territories not affected directly by the influential players. 

Our numerical investigations have clearly shown that the temporary links between the moving influential players promote the spreading (and maintenance) of cooperative behavior. In comparison with the case of quenched distribution of $A$ players, the quantitative analysis has confirmed that higher level of cooperation can be achieved if the system has less number of influential players who can move randomly with an optimal rate.

\begin{acknowledgments}

This work was supported by the Hungarian National Research Fund
(Grant No. 73449) and by the COST P10 project ''Physics of Risk''.

\end{acknowledgments}

\bibliography{egg}

\end{document}